\documentclass[usenatbib]{mn2e}
\usepackage{psfig}

\def\gs{\mathrel{\raise0.35ex\hbox{$\scriptstyle >$}\kern-0.6em\lower0.40ex\hbox{{$\scriptstyle \sim$}}}}
\def\ls{\mathrel{\raise0.35ex\hbox{$\scriptstyle <$}\kern-0.6em\lower0.40ex\hbox{{$\scriptstyle \sim$}}}}

\def\Wm2{\,\hbox{W}\,\hbox{m}^{-2}}
\def\gsim{\mathrel{\raise0.35ex\hbox{$\scriptstyle >$}\kern-0.6em\lower0.40ex\hbox{{$\scriptstyle \sim$}}}}
\def\lsim{\mathrel{\raise0.35ex\hbox{$\scriptstyle <$}\kern-0.6em\lower0.40ex\hbox{{$\scriptstyle \sim$}}}}

\begin{document}

\title[HST imaging of submillimetre galaxies]{A Hubble Space Telescope
  NICMOS and ACS Morphological Study of $z\sim$2 Submillimetre
  Galaxies}

\author[Swinbank et al.]
{\parbox[h]{\textwidth}{ 
A.\,M.\ Swinbank$^{\, 1,*}$,
Ian Smail$^{\, 1}$,
S.\,C.\ Chapman$^{\, 3}$,
C.\ Borys$^{\, 2}$, \\
D.\,M.\ Alexander$^{1}$,
A.\,W.\ Blain$^{\, 2}$,
C.\,J.\ Conselice$^{\, 4}$,
L.\,J.\ Hainline$^{\, 5}$,
\& R.\,J.\ Ivison$^{\, 6,7}$,
}
\vspace*{6pt} \\
$^1$Institute for Computational Cosmology, Department of Physics, Durham University, South Road, Durham, DH1 3LE, UK \\
$^2$IPAC, California Institute of Technology, 100-22, Pasadana CA 91125 USA
$^3$Institute of Astronomy, University of Cambridge, Madingley Road, Cambridge, CB3 0HA, UK \\
$^4$School of Physics and  Astronomy, Nottingham University, University Park, Nottingham,  NG7 2RD, UK \\
$^5$Department of Astronomy, University of Maryland, College Park, MD 20742, USA \\
$^6$UK Astronomy Technology Centre, Royal Observatory, Blackford Hill, Edinburgh, EH19 3HJ, UK \\
$^7$Institute for Astronomy, University of Edinburgh, Edinburgh, EH19 3HJ, UK \\
$^*$Email: a.m.swinbank@durham.ac.uk \\
}

\maketitle

\begin{abstract}
  We present a quantitative morphological analysis using {\it Hubble
    Space Telescope} {\it (HST)} NICMOS $H_{160}$- and ACS
  $I_{775}$-band imaging of 25 spectroscopically confirmed
  submillimetre galaxies (SMGs) which have redshifts between
  $z=0.7$--3.4 ($\bar{z}=2.1$).  Our analysis also employs a comparison
  sample of more typical star-forming galaxies at similar redshifts
  (such as Lyman Break Galaxies) which have lower far-infrared
  luminosities.  This is the first large-scale study of the
  morphologies of SMGs in the near-infrared at $\sim 0.1''$ resolution
  ($\lsim$\,1\,kpc).  We find that the half light radii of the SMGs
  (r$_{h}$=2.3$\pm$0.3 and 2.8$\pm$0.4\,kpc in the observed $I$- and
  $H$-bands respectively) and asymmetries are not statistically
  distinct from the comparison sample of star-forming galaxies.
  However, we demonstrate that the SMG morphologies differ more {\it
    between} the rest-frame UV and optical-bands than typical
  star-forming galaxies and interpret this as evidence for structured
  dust obscuration.  We show that the composite observed $H$-band light
  profile of SMGs is better fit with a high Sersic index ($n\sim2$)
  than with an exponential disk suggesting the stellar structure of
  SMGs is best described by a spheroid/elliptical galaxy light
  distribution.  We also compare the sizes and stellar masses of SMGs
  to local and high-redshift populations, and find that the SMGs have
  stellar densities which are comparable (or slightly larger) than
  local early-type galaxies, but comparable to luminous, red and dense
  galaxies at $z\sim$1.5 which have been proposed as direct SMG
  descendants, although the SMG stellar masses and sizes are
  systematically larger.  Overall, our results suggest that the
  physical processes occuring within the galaxies are too complex to be
  simply characterised by the rest-frame UV/optical morphologies which
  appear to be essentially decoupled from all other observables, such
  as bolometric luminosity, stellar or dynamical mass.
\end{abstract}


\section{Introduction}
\label{secintro}

Around 60\% of the stellar mass in the local Universe is contained
within early-type and elliptical galaxies, which sit on a tight ``red
sequence'' in the colour magnitude diagram
\citep{Sandage78,Bower92,Bell03}.  Early-type and elliptical galaxies
follow well known scaling relations (the fundamental plane) and exhibit
systematic correlations between the absorption line strengths and
velocity dispersion, $(\sigma)$.  This age-$\sigma$ relation is such
that the most massive ($\sigma\sim$400\,km\,s$^{-1}$) galaxies appear
to have formed their stars $\sim$10--13 Gyr ago.  By contrast the mean
age of the lower dispersion galaxies ($\sigma\sim$50\,km\,s$^{-1}$)
formed more recently; $\sim$4\,Gyr ago \citep[e.g.][]{Smith07}.  This
suggests that the stars in giant red galaxies formed early in the
history of the Universe \citep[eg.][]{Nelan05}.  Using deep
near-infrared imaging, it has also become possible to extend the
selection of red galaxies to higher redshift and identify massive,
relatively old galaxies at $z\sim1.5$ which could be considered the
progenitors of the local elliptical population
(eg. \citealt{vanDokkum04,Cimatti08}).  Clearly to probe this
evolutionary sequence further, direct observations of the formation of
the most massive galaxies at high-redshift are required.  However, this
has turned out to be a non-trivial exercise since the most
actively star-forming, massive galaxies at $z>$2 are also the most dust
obscured \citep{Dole04,Papovich04,LeFloch09}.

Nevertheless, mid- and far- infrared surveys (particularly those made
with the 850$\mu$m SCUBA camera on the JCMT and more recently with the
{\it Spitzer Space Telescope} at 24$\mu$m) have begun to resolve the
most highly obscured populations into their constituent galaxies,
determine their contribution to the energy density in the
extra-galactic far-infrared/sub-mm background, and chart that history
of massive galaxy formation \citep{Smail02,Cowie02,LeFloch05}.
Extensive, multi-wavelength follow-up has shown that these heavily
dust-obscured, gas-rich galaxies lie predominantly at high redshift
($z\sim2$; e.g. \citealt{Chapman03a,Chapman05a}), with bolometric
luminosities of $\gg$10$^{12}$L$_{\odot}$ and star-formation rates of
order 700\,M$_{\odot}$\,yr$^{-1}$.  It has therefore been speculated
that SMGs are the progenitors of luminous elliptical galaxies
(e.g.\ \citealt{Lilly99,Genzel03,Blain04a,Swinbank06b,Tacconi08}).

With the redshift distributions and contributions to the cosmic energy
density of ultra-luminous galaxies reasonably constrained, the next
step is to study the evolutionary history of SMGs, and to determine how
they relate to lower luminosity galaxies.  Indeed, given the apparently
rapid evolution in the space density of ULIRGs from $z\sim2$ to $z=0$
\citep{Chapman05a,LeFloch05}, one key issue is to understand the
physical processes which trigger these far-infrared luminous events.
Indeed, the mechanism responsible for these vigorous starbursts is
still uncertain.  Analogy to local ULIRGs would argue for merging as
the trigger, although secular bursts in massive gas disks is also
conceivable and indeed recent theoretical interest has stressed the
importance of cold flows in high-redshift star formation
\citep[e.g.][]{Genel08}. The suggestion that SMGs have compact
disk-like gas reservoirs ($R_{1/2}<2$\,kpc) with ``maximal starbursts''
\citep{Tacconi08} hints that SMGs are scaled-up versions of the local
ultra-luminous galaxy population, which are usually associated with
merger activity \citep{Tacconi02}.

In order to test the connection between SMGs, lower-luminosity
star-forming galaxies at high redshift, as well as local ULIRGs, we
have obtained high resolution {\it Hubble Space Telescope (HST)}
imaging of a sample of spectroscopically confirmed SMGs at
$z=$0.7--3.4.  By necessity, most morphological studies of high
redshift galaxies to date have been performed at optical wavelengths
which probe the rest-frame UV
\citep{Chapman03b,Webb03,Conselice03b,Conselice08,Smail04,Law07b}.
However the rest-frame UV is dominated by radiation which traces the
brightest, active star-forming regions rather than the bulk of the
stellar population and can lead to late-type galaxies being classified
as irregular systems \citep{Dickinson00,Thompson03,Goldader02}.
Nevertheless, in a recent study, \citet{Law07b} conducted a detailed
analysis of the rest-frame UV morphologies of a large sample of
UV/optically selected star-forming galaxies at $z\sim$1--3 in the
Hubble Deep Field North (HDFN) and find evidence that dusty galaxies
have more nebulous UV morphologies than more typical sources, but
otherwise conclude that UV morphology is statistically decoupled from
the majority of physical observables (such as stellar or dynamical
mass, gas fraction or star-formation rate).  Here we aim to extend this
work to include the rest-frame optical emission to test whether there
are key differences in the morphologies at longer wavelengths (as
suggested by imaging of low redshift ULIRGs;
e.g. \citealt{Goldader02}).  We have therefore assembled a sample of 25
SMGs with both {\it HST} ACS $I$-band and NICMOS $H$-band observations.
We determine the basic morphological parameters of this sample of SMGs,
as well as search for signs of tidal features and major mergers.  To
baseline our analysis we also use a sample of 228 optically selected
star-forming galaxies at $z\sim2$--3 (of which 53 have also been
observed in $H$-band with {\it HST}).

We use a {\it WMAP} cosmology \citep{Spergel04} with
$\Omega_{\Lambda}$=0.73, $\Omega_{m}$=0.27, and
H$_{0}$=72\,km\,s$^{-1}$\,Mpc$^{-1}$.  In this cosmology, at $z=$2.1
(the median redshift of our SMG sample), 0.1$"$ (the typical resolution
of our observations) corresponds to a physical scale of 0.8\,kpc.  All
quoted magnitudes are on the AB system unless otherwise noted.

\begin{figure*}
\centerline{\psfig{file=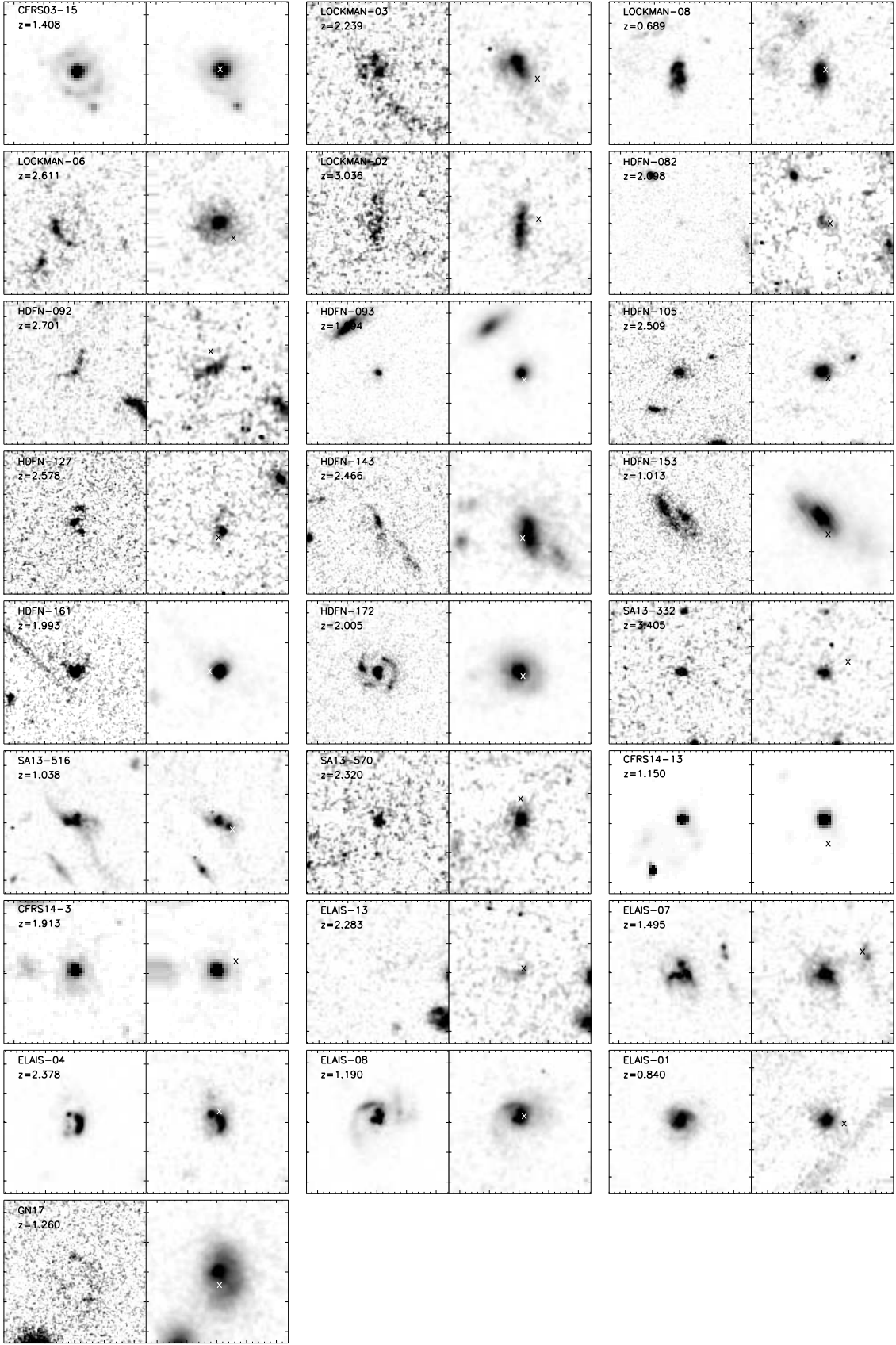,angle=0,width=6.0in}}
\caption{High resolution {\it HST} optical (ACS/WFPC2/STIS) and
  near-infrared NICMOS imaging of SMGs.  For each galaxy, the left hand
  panel denotes the optical image whilst the right hand panel denotes
  the 1.6$\mu$m ($H-$band) image.  Each thumbnail is scaled such that
  they each cover 40 kpc at the redshift of the SMG.  The small cross
  denotes the location of the centroid of the radio emission.}
\label{fig:hstthumbs1}
\end{figure*}

\begin{figure*}
\centerline{\psfig{file=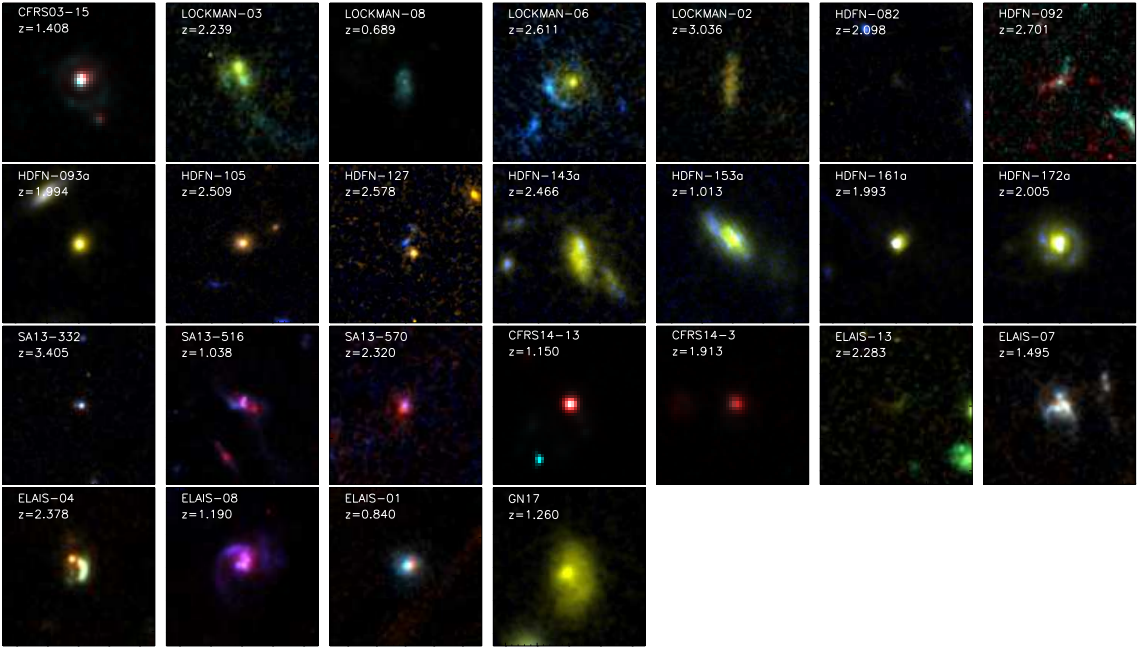,angle=0,width=6.0in}}
\caption{True colour {\it HST} $IH$-band images of the SMGs in our
  sample showing the range of colours and morphological mix within the
  sample.  Each image is 40\,kpc at the redshift of the galaxy.}
\label{fig:colSMG}
\end{figure*}

\begin{figure*}
\centerline{\psfig{file=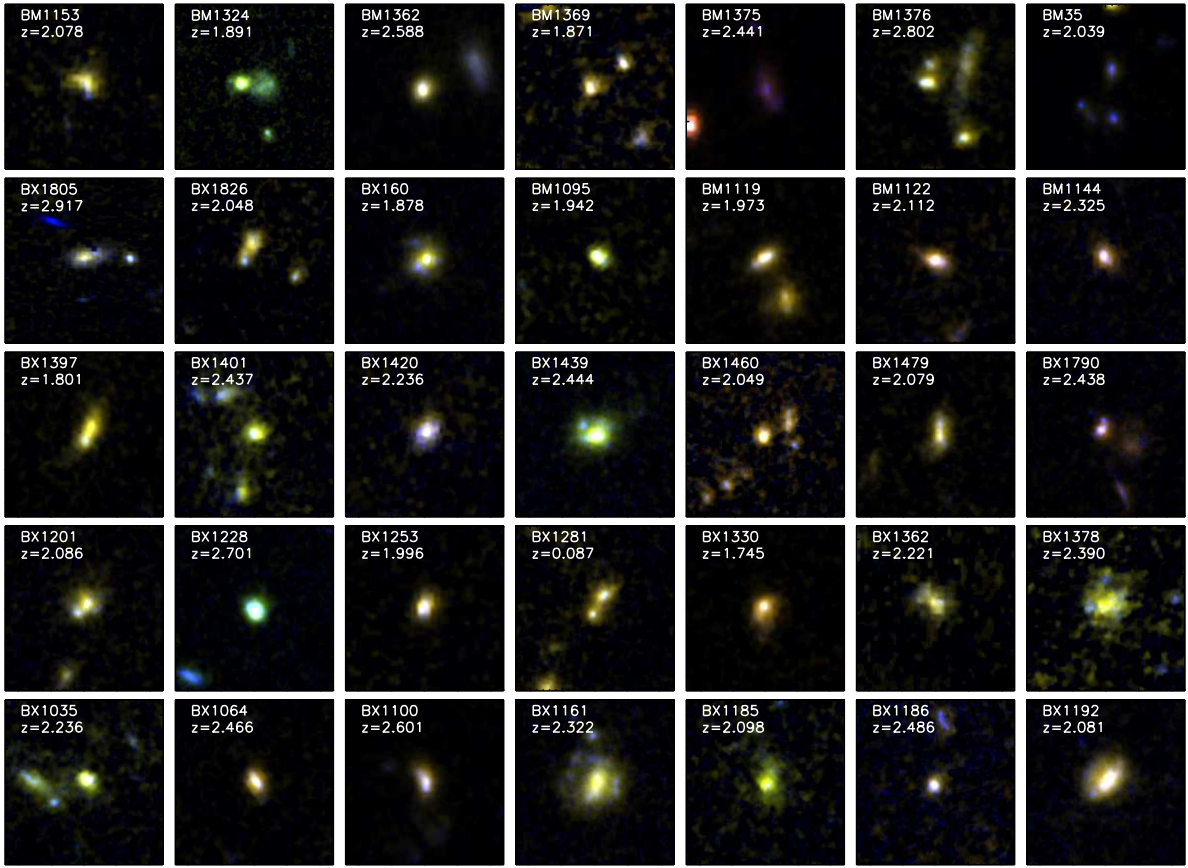,angle=90,width=6.4in}}
\caption{Comparison true colour {\it HST} $IH$-band images of thirty
  star-forming galaxies at $z\sim2$ from the spectroscopic sample of
  \citet{Reddy06b} showing the mix of morphological types is comparable
  to the SMGs.  As in Fig.~\ref{fig:hstthumbs1}, each image is 40\,kpc
  at the redshift of the galaxy so that a direct comparison can me made.}
\label{fig:colLBG}
\end{figure*}

\begin{figure*}
\centerline{
\psfig{file=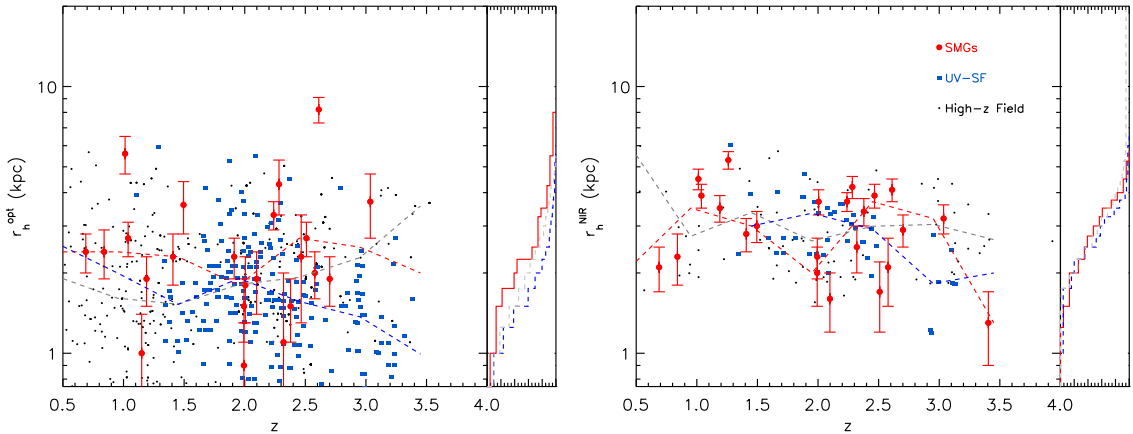,width=6in,angle=90}}
\caption{Size versus redshift relation for SMGs (filled red circles)
  compared to the UV-SF (filled blue square) and high-$z$ field (small
  black squares) comparison samples.  The half light radii in the left
  hand panel are derived using {\it HST} $I$-band data whilst the right
  hand panel are derived from the NICMOS $H$-band (F160W) imaging.  The
  dashed lines show that the medians of the distribution and illustrate
  that the SMGs and comparison samples have comparable half light radii
  in both the $I$- and $H$-bands.  On the right hand side of each
  sub-panel we show the cumulative histograms of each distribution
  which shows that in the rest-frame UV the SMGs have marginally larger
  half light radii on average than the comparison samples, but
  essentially indistinguishable distributions in the rest-frame
  optical.}
\label{fig:rh_z}
\end{figure*}

\begin{figure}
\centerline{
\psfig{file=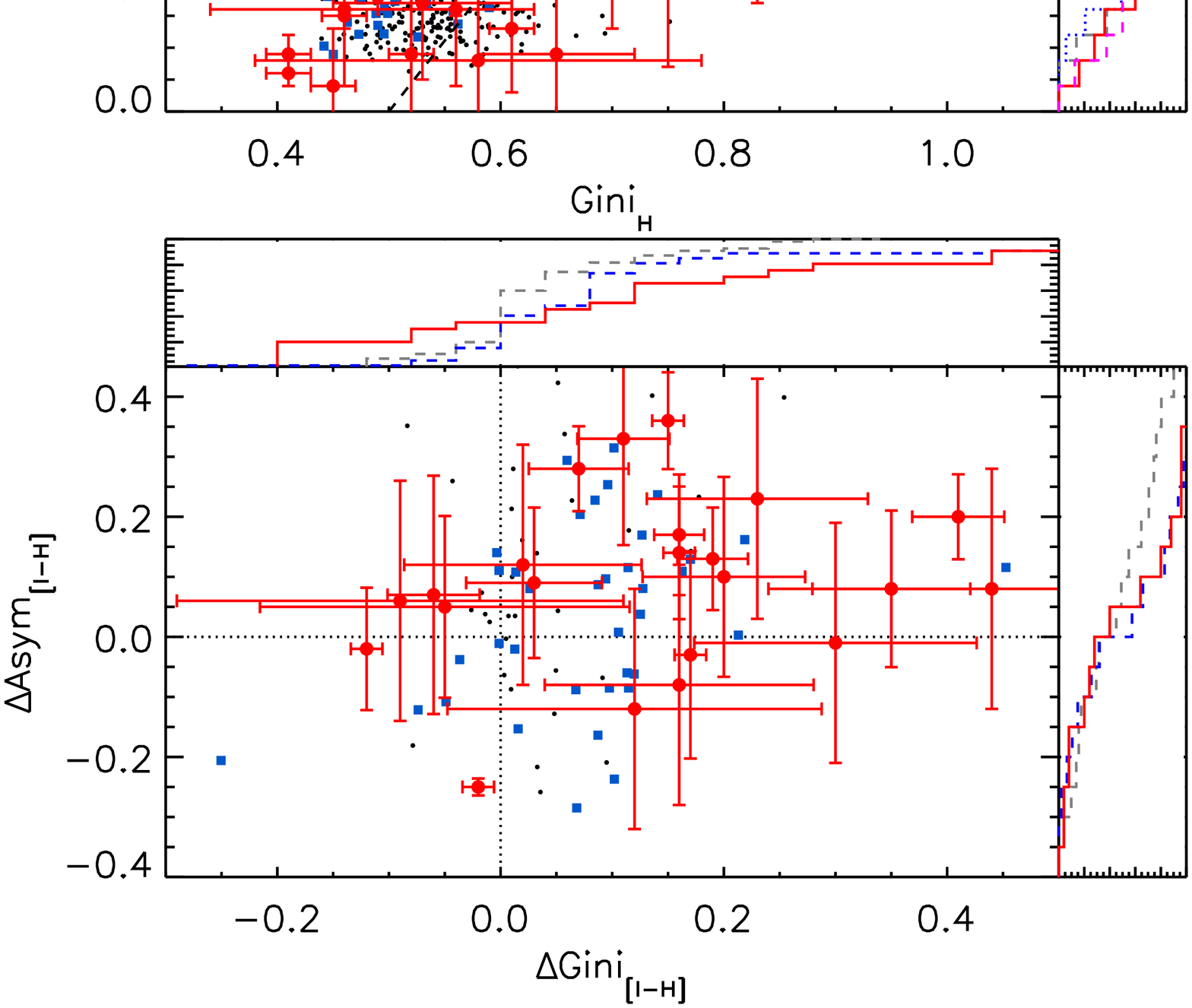,angle=90,width=3.0in}}
\caption{Comparison between the structural parameters for SMGs compared
  to star-forming field galaxies.  {\it Top:} Gini versus Asymmetry
  measured in the observed $I$-band for the SMGs showing that
  the median asymmetry for the SMGs and comparison samples is
  comparable, but that there is an offset between the Gini
  co-efficients of $\Delta G\sim$0.1 between SMGs and the comparison
  samples.  The dashed line illustrates the correlation between Gini
  and Asymmetry in local galaxies (E-Sd morphological types) from Lotz
  et al. (2004).  The cumulative histograms on each axis show that
  there is a significant difference in the rest-frame UV Gini
  coefficient for the SMGs compared to the comparison samples, but the
  asymmetries are comparable to the comparison samples. {\it Centre:}
  $H$-band Gini versus Asymmetry for the SMGs and field star-forming
  galaxies.  As in the observed $I$-band the asymmetry is
  indistinguishable between the populations.  However, the median Gini
  coefficient for the SMGs in only $\sim$0.03 larger than the field
  population. The dashed line denotes the same correlation as shown in
  the top panel.  The cumulative histograms show that there is a subtle
  difference between the Gini co-efficient in the rest-frame optical
  for the SMGs and comparison samples, but the asymmetries are
  indistinguishable.  {\it Bottom:} $\Delta G$ versus $\Delta A$ for the
  SMGs and field galaxies showing that the both the SMGs and field
  populations tend to prefer larger $\Delta G$ values.}
\label{fig:GiniAsym}
\end{figure}

\section{Observations, Reduction and Analysis}

\subsection{SMG Sample}

We used {\it HST} NICMOS during Cycle 12 (PID: 9506) to observe 23
galaxies spanning the redshift range $0.7<z<3.4$ from the spectroscopic
catalog of \citet{Chapman05a}.  We also include in our analysis two
spectroscopically confirmed SMGs from \citet{Borys04} (see also
\citealt{Pope05}) which are part of the \citet{Chapman05a} sample and
which lie in the GOODS-North (GOODS-N) field and were observed with
NICMOS during program PID: 11082 (Conselice et al.\ 2010 in prep).
Therefore the final sample consists of 25 SMGs with secure
spectroscopic redshifts, and localized via VLA radio imaging.

The majority of the NICMOS targets also have optical coverage from
\textsl{HST} as part of the same program from Cycle 12, although we
also include archival data in the ELAIS field (PID: 9761) and GOODS
Legacy program.  Briefly, 20 galaxies have \textsl{HST} ACS imaging,
four galaxies have WFPC2 observations, and one galaxy has STIS imaging.
Unless otherwise stated, we will refer to the ACS, WFPC2, and STIS
images as ``optical'', and specifically the $I$-band filter.  In the
case of the lone STIS image, no filter was used (hence the wavelength
range covered is $\sim$5400--1$\mu$m).  We note that the sample spans a
redshift range $z$=0.7--3.4 (median redshift $z=$2.1), 850$\mu$m flux
($S_{850}=$3--15\,mJy) and $I$-band magnitude range ($I$=21.5--26.5)
which is representative of the parent sample of 73 SMGs in
\citet{Chapman05a}.  We also note that there are roughly equal numbers
for each $\Delta{z}=1$ bin (Table~1).

The near-infrared observations were made using the NIC2 camera and the
F160W filter.  We employed the standard spiral dither pattern under
LOWSKY conditions.  Each exposure was corrected for a pedestal offset,
and then mosaiced together using the {\sc calnicb} task in {\sc IRAF}.
Unfortunately several exposures were affected by the South Atlantic
Anomaly (SAA), and extra processing steps were
required\footnote{\texttt{http://www.stsci.edu/hst/nicmos/tools/post\_SAA\_tools.html}}.
The final images appear very flat and have very low cosmic ray
contamination.  The observed $I$ and $H$-band images of each SMG in
this sample are shown in Fig.~\ref{fig:hstthumbs1}.


\subsection{Comparison Sample}

To construct a comparison sample, we exploit the extensive spectroscopy
in GOODS-N.  To act as a high redshift ``field'' sample, we use
$\sim$2100 galaxies from \citet{Barger08} between $z=1$--3.5 of which
330 also lie within the GOODS-N NICMOS imaging (Conselice et al. 2010
in prep).  To ensure a fair comparison, we restrict this high-redshift
field sample to have the same redshift distribution as the SMGs (we
note that the final high-redshift field sample has an $I$-band
magnitude distribution of $I$=22.78$\pm$1.05; for reference, the SMGs
have $I$=23.37$\pm$1.6).  We hereafter refer to this sample as the
``high-$z$ field'', but caution that the selection function for this
sample is extremely complex since it comprises a highly incomplete mix
of UV/optical, mid-infrared and X-ray selected galaxies.  We therefore
also restrict the analysis specifically to a more homogeneously
selected sample of UV/optically selected galaxies (star-forming BX/BM
and LBGs) with secure spectroscopic redshifts (in the range
$z\sim$1.6--3.5) from \citet{Reddy06b} (hereafter called UV-SF
galaxies).  This sample contains 228 galaxies which are in the
GOODS-North ACS $I$-band imaging of which 53 are also have also been
observed with NICMOS.  We also remove from the UV-SF sample one galaxy
which has a bolometric luminosity greater than $10^{12}$L$_{\odot}$ and
note that the median bolometric luminosity of the remaining sample is
L$_{bol}\sim$10$^{11.4}L_{\odot}$ \citep{Reddy06b} which is an order of
magnitude lower than the typical bolometric luminosity of the SMGs.  In
Fig.~\ref{fig:colSMG} and ~\ref{fig:colLBG} we show true-colour {\it
  HST} images of the SMGs and UV-SF galaxies to demonstrate the visual
mix of morphological types is comparable.

\subsection{Galaxy Sizes and Morphologies}

To quantify the galaxy morphologies (and measure their physical scale),
we first calculate the Petrosian and half-light radii in the optical
and near-infrared.  The Petrosian radius is defined by
r$_{pet}$=1.5$\times$r$_{\eta=0.2}$ where $\eta=0.2$ is the radius
($r$) at which the surface brightness within an annulus at $r$ is one
fifth of the surface brightness within $r$
\citep[][]{Conselice00,Chapman03b}.  This provides a measure of the
size which does not depend on isophotes.  The half-light radius,
$r_{h}$, is then defined as the radius at which the flux is one-half of
that within $r_{pet}$.

Given the small angular sizes of these objects, and their apparently
complex morphologies, it is difficult to apply a standard morphological
analysis along the lines of a ``Hubble tuning fork''.  Fortunately,
statistics have been developed to help characterize high redshift
galaxy morphologies.  We concentrate on the Gini and Asymmetries
\citep{Abraham03,Conselice03c} although we note that a number of other
parameterisations of galaxy morphologies have been developed (such as
concentration and clumpiness as part of the CAS system;
\citealt{Conselice03c}).  However, since the concentration parameter
represents the scale of the galaxy, and the clumpiness defines the
spatial light distribution, to keep the analysis concise here we
concentrate on the half-light radius and Gini co-efficient which are
similar parameterisations.

The Gini coefficient is a statistical tool, originally developed for
economics, which determines the distribution of wealth within a
population.  Higher values indicate a very unequal distribution ($G$=1
indicates all of the flux is in one pixel), whilst a lower value
indicates it is more widely distributed ($G$=0 suggests the flux is
evenly distributed).  The value of $G$ is defined by the Lorentz curve
of the galaxies light distribution, which does not take into
consideration any of the spatial information.  The value $G$ is derived
by first sorting the pixel values within the Petrosian aperture, and
then summed over a cumulative distribution (see \citealt{Abraham03}).

Since the Gini coefficient removes all spatial information, to study
the structures and morphologies of our sample, we also compute the
galaxy asymmetry ($A$).  The asymmetry of a galaxy is measured by
minimising the subtraction of a galaxy image from itself after rotating
the original image by 180$^{\circ}$.  The minimisation takes into
account the uncertainty in deriving the galaxy center (see
\citet{Conselice08} and references therein for a detailed discussion).
Briefly, a lower value of $A$ suggests that the galaxy is more
symmetrical (e.g.\, elliptical galaxies), whilst higher values of $A$
suggest highly asymmetric systems, usually found in spiral galaxies or
major mergers (for reference, in the rest-frame UV/optical, an
asymmetry value in excess of A=0.30 has been found to be a robust
indication of a major merger at $z$=0 \citealt{Conselice03a})

To derive the asymmetry for the galaxies in our sample, we first
extract a 10$''\times$10$''$ thumbnail of a galaxy image from the
background subtracted frame.  We mask all objects except for the galaxy
image using the sextractor segmentation map, replacing these regions
with values taken from the correct noise properties as measured from
the sky.  We then measure the petrosian radius, half light radius and
Gini co-efficient ($r_{pet}, r_h$ \& $G$).  To derive the asymmetry, we
first derive a crude centre for the galaxy by finding the luminosity
weighted centroid of the galaxy light distribution.  We extract, rotate
and subtract the galaxy image about this center, but then iterate and
allow the center to vary over the entire half-light radius.  This
function is minimised to provide the integer pixel centre of the
galaxy.  Operationally, after the initial centre is found, the
asymmetry is computed again for centres at the surrounding eight points
in a 3$\times$3 grid.  We use a distance of 0.1 pixels, corresponding
to approximately 0.1\% of the galaxy half light radius and use bilinear
interpolation on the shifted image.  If the asymmetry parameter at the
centre is lower than any of the surrounding pixels, then the asymmetry
parameter is taken as this value.  If the central pixel does not give
the asymmetry minimum, then the procedure is repeated with the new
centre where the minimum was found.  This process repeats until the
minimum is found.

\section{Analysis and Results}

\subsection{Galaxy Sizes}

First, we compare the half light sizes of the SMGs and comparison
samples.  As Fig.~\ref{fig:rh_z} shows, the SMGs 
have an optical median half light radii of $r_{h}^{opt}({\rm SMG})$=2.3$\pm$0.3\,kpc, 
which significantly overlaps with the half-light radii of UV-SF
galaxies ($r_{h}^{opt}({\rm UV-SF})$=1.9$\pm$0.2\,kpc),
and the field population ($r_{h}^{opt}({\rm high-z\,
  field})$=2.0$\pm$0.1\,kpc) (here we quote the error on the mean and
we note that in all cases the 1$\sigma$ scatter in the distribution is
is 1.3--1.7\,kpc).  These rest-frame UV SMG half light radii are
similar to those found by \citet{Chapman03b} who derive
r$_h$=2.8$\pm$0.3\,kpc for a small sample of SMGs.

Turning to the observed $H$-band, the SMGs have a median half light
radii of $r_{h}^{nir}({\rm SMG})$=2.8$\pm$0.4\,kpc which is in good
agreement with both the high-$z$ field population, and the UV-SF sample
which have $r_{h}^{nir}$=2.6$\pm$0.2 and 2.5$\pm$0.2\,kpc respectively.
Thus it appears that the sizes of the SMGs are not systematically
larger than the lower-luminosity star-forming galaxies at the same
epoch in either $I$- or $H$-bands (Table~2), but the near-infrared half
light radii tend to be systematically larger ($\sim$0.5\,kpc) in all
three galaxy populations.  Finally, in Fig.~\ref{fig:rh_z} we show the
optical and near-infrared sizes for all three galaxy samples as a
function of redshift from $z=$0.5--3.5.  Binning the samples into
$\Delta z$=0.5 bins, none of the three samples show systematic trends
of half light radius with redshift.

%
%
%
\begin{table*}
\begin{center}
{\footnotesize
{\centerline{\sc Table 1.}}
{\centerline{\sc Log of SMG sample with NICMOS imaging}}
\begin{tabular}{lcccccccc}
\hline
\noalign{\smallskip}
ID                                       & Short Name  & $z$    & Type     & F160W            &  $r^{nir}_{pet}$  &  $r^{nir}_{h}$ & $r^{opt}_{pet}$ & $r^{opt}_{h}$ \\
                                         &             &        &          & (AB)             &   (kpc)          &  (kpc)       &  (kpc)         &  (kpc) \\
\hline
  SMM\,J030227.73+000653.5$^{1}$         & CFRS03-15   & 1.408  & SB       & $20.73\pm0.02$   & 14.4$\pm$2.0  &  2.8$\pm$0.4 & 13.1$\pm$1.5  &  2.3$\pm$0.5 \\ 
  SMM\,J105158.02+571800.2               & LOCKMAN-03  & 2.239  & SB       & $21.58\pm0.04$   &  9.9$\pm$1.3  &  3.7$\pm$0.3 &  7.8$\pm$1.3  &  3.3$\pm$0.4 \\ 
  SMM\,J105200.22+572420.2               & LOCKMAN-08  & 0.689  & SB       & $22.98\pm0.08$   &  5.3$\pm$0.5  &  2.1$\pm$0.4 &  5.0$\pm$1.3  &  2.4$\pm$0.4 \\ 
  SMM\,J105230.73+572209.5               & LOCKMAN-06  & 2.611  & SB       & $22.11\pm0.05$   & 10.2$\pm$0.5  &  4.1$\pm$0.4 & 15.9$\pm$2.5  &  8.2$\pm$0.9 \\ 
  SMM\,J105238.30+572435.8               & LOCKMAN-02  & 3.036  & SB       & $22.43\pm0.08$   &  8.6$\pm$0.6  &  3.2$\pm$0.4 &  5.7$\pm$2.0  &  3.7$\pm$1.0 \\ 
  SMM\,J123553.26+621337.7               & HDFN-082    & 2.098  & SB       & $24.30\pm0.21$   &  3.1$\pm$0.2  &  1.6$\pm$0.4 &  3.7$\pm$0.4  &  1.9$\pm$0.5 \\ 
  SMM\,J123600.10+620253.5$^{2}$         & HDFN-092    & 2.701  & SB       & $23.75\pm0.16$   &  5.0$\pm$1.5  &  2.9$\pm$0.4 &  3.6$\pm$1.2  &  2.4$\pm$0.4 \\ 
  SMM\,J123600.15+621047.2               & HDFN-093    & 1.994  & SB       & $22.64\pm0.08$   &  4.2$\pm$2.0  &  2.3$\pm$0.4 &  5.0$\pm$0.8  &  1.5$\pm$0.4 \\ 
  SMM\,J123606.85+621021.4               & HDFN-105    & 2.509  & SB       & $21.61\pm0.03$   &  4.2$\pm$0.5  &  1.7$\pm$0.5 &  7.6$\pm$0.5  &  2.7$\pm$0.4 \\ 
  SMM\,J123616.15+621513.7               & HDFN-127    & 2.578  & SB       & $23.59\pm0.10$   &  9.7$\pm$2.0  &  2.1$\pm$0.6 &  7.6$\pm$1.2  &  2.0$\pm$0.4 \\ 
  SMM\,J123622.65+621629.7               & HDFN-143    & 2.466  & SB       & $23.32\pm0.09$   &  9.7$\pm$0.4  &  3.9$\pm$0.4 &  6.1$\pm$2.0  &  2.3$\pm$1.0 \\ 
  SMM\,J123629.13+621045.8               & HDFN-153    & 1.013  & SB       & $21.23\pm0.03$   & 12.0$\pm$0.9  &  4.5$\pm$0.4 & 13.8$\pm$2.5  &  5.6$\pm$0.9 \\ 
  SMM\,J123632.61+620800.1               & HDFN-161    & 1.993  & AGN      & $22.74\pm0.07$   &  5.0$\pm$0.4  &  2.0$\pm$0.5 &  2.5$\pm$0.5  &  0.9$\pm$0.4 \\ 
  SMM\,J123635.59+621424.1               & HDFN-172    & 2.005  & AGN      & $21.59\pm0.03$   &  9.2$\pm$0.4  &  3.7$\pm$0.4 & 10.6$\pm$1.2  &  1.8$\pm$0.5 \\ 
  SMM\,J123701.59+621513.9               & GN17        & 1.260  & SB       & $21.50\pm0.05$   & 12.5$\pm$0.4  &  5.3$\pm$0.4 &    ....       &  ....        \\ 
  SMM\,J131201.17+424208.1               & SA13-332    & 3.405  & AGN      & $22.87\pm0.08$   &  3.3$\pm$0.4  &  1.3$\pm$0.4 &  2.2$\pm$0.4  &  0.4$\pm$0.4 \\ 
  SMM\,J131225.20+424344.5$^{1}$         & SA13-516    & 1.038  & SB       & $21.46\pm0.03$   & 10.3$\pm$1.2  &  3.9$\pm$0.4 &  8.9$\pm$0.5  &  2.7$\pm$0.4 \\ 
  SMM\,J131232.31+423949.5               & SA13-570    & 2.320  & SB       & $22.85\pm0.09$   &  7.5$\pm$0.9  &  2.5$\pm$0.5 &  3.6$\pm$1.5  &  1.1$\pm$1.0 \\ 
  SMM\,J141741.81+522823.0$^{1}$         & CFRS14-13   & 1.150  & AGN      & $18.67\pm0.02$   &  5.6$\pm$1.5  &  $<1$        &  6.6$\pm$2.0  &  1.0$\pm$0.4 \\ 
  SMM\,J141800.40+522820.3$^{1}$         & CFRS14-3    & 1.913  & SB       & $22.74\pm0.07$   &  3.7$\pm$0.6  &  $<1$        &  8.3$\pm$0.3  &  2.3$\pm$0.4 \\ 
  SMM\,J163631.47+405546.9              & ELAIS-13    & 2.283  & AGN      & $24.47\pm0.18$   &  8.6$\pm$0.6  &  4.2$\pm$0.4 &  7.3$\pm$2.0  &  4.3$\pm$1.0 \\ 
  SMM\,J163639.01+405635.9              & ELAIS-07    & 1.495  & SB       & $23.88\pm0.09$   &  8.3$\pm$0.7  &  3.0$\pm$0.4 & 12.7$\pm$3.0  &  3.6$\pm$0.8 \\ 
  SMM\,J163650.43+405734.5              & ELAIS-04    & 2.378  & SB/AGN   & $21.85\pm0.04$   &  9.7$\pm$0.6  &  3.4$\pm$0.4 &  5.0$\pm$1.5  &  1.5$\pm$0.4 \\ 
  SMM\,J163658.78+405728.1              & ELAIS-08    & 1.190  & SB       & $21.20\pm0.03$   &  9.9$\pm$0.6  &  3.5$\pm$0.4 &  6.8$\pm$2.0  &  1.9$\pm$0.4 \\ 
  SMM\,J163704.34+410530.3              & ELAIS-01    & 0.840  & SB       & $22.36\pm0.03$   &  6.9$\pm$1.2  &  2.3$\pm$0.5 &  6.1$\pm$0.5  &  2.4$\pm$0.5 \\ 
\hline
\label{table:phot}
\end{tabular}
}
\caption{Notes: The ID's are taken from Chapman et al.\ 2005.  ID's
  marked with a $^{1}$ denote galaxies observed with WFPC2 in the
  $I$-band whilst the galaxy marked by $^{2}$ denotes observations with
  STIS.  The starburst (SB) versus AGN classification is taken from
  optical and near-infrared spectroscopy from \citet{Chapman05a} and
  \citet{Swinbank04}.  $r_{pet}$ and r$_{h}$ denote pertrosian and half
  light radii respectively.  SMM\,J123635.59+621424.1 (GN17) is
  optically faint and so we have not attempted to derive petrosian and
  half light radii.}
\end{center}
\end{table*}

%
%
%
\begin{table*}
\begin{center}
{\footnotesize
{\centerline{\sc Table 2.}}
{\centerline{\sc Median Morphological Parameters}}
\begin{tabular}{lcccccccccc}
\hline
\noalign{\smallskip}
Sample & r$^{opt}_{pet}$(kpc) & r$^{nir}_{pet}$(kpc)  & r$^{opt}_h$(kpc)  & r$^{nir}_h$(kpc)  & G$^{opt}$     & G$^{nir}$      & A$^{opt}$     & A$^{nir}$      & $\Delta G$    & $\Delta A$    \\
\hline
SMGs  & 6.9$\pm$0.7 & 7.7$\pm$0.6 & 2.3$\pm$0.3 & 2.8$\pm$0.4 & 0.69$\pm$0.03 & 0.56$\pm$0.02 & 0.27$\pm$0.03 & 0.25$\pm$0.02 & 0.13$\pm$0.04 & 0.02$\pm$0.03 \\
UV-SF & 6.2$\pm$0.2 & 6.6$\pm$0.3 & 1.9$\pm$0.2 & 2.6$\pm$0.2 & 0.59$\pm$0.02 & 0.53$\pm$0.02 & 0.29$\pm$0.02 & 0.24$\pm$0.02 & 0.06$\pm$0.03 & 0.05$\pm$0.03 \\
High-$z$ field & 7.7$\pm$0.2 & 7.8$\pm$0.2 & 2.0$\pm$0.1 & 2.5$\pm$0.2 & 0.60$\pm$0.05 & 0.55$\pm$0.01 & 0.24$\pm$0.10 & 0.24$\pm$0.01 & 0.05$\pm$0.05 & 0.00$\pm$0.10 \\
\hline
\label{table:morph}
\end{tabular}
}
\end{center}
\end{table*}

%
%
%

\subsection{Gini and Asymmetries}

In Fig.~\ref{fig:GiniAsym} we show the $I$- and $H$-band Gini and
Asymmetry coefficients for the SMGs compared to the UV-SF sample and
the high-$z$ field population.  In the observed $I$-band the SMGs have
a median Gini value of $G^{opt}({\rm
  SMG})$=0.69$\pm$0.03. 
In contrast, the comparison samples have $G^{opt}({\rm
  UV-SF})$=0.59$\pm$0.02 and $G^{opt}({\rm high-z\,
  field})$=0.60$\pm$0.05.  Thus the SMGs have a systematically larger
optical Gini coefficient ($\Delta G\sim$0.1) than the comparison
samples.  In the observed $H$-band, the SMGs also have a slightly
larger Gini co-efficient with $G^{nir}({\rm SMG})$=0.56$\pm$0.02
which is $\Delta G=0.03$ larger than the UV-SF population.  The
differences in the Gini co-efficient between SMGs and UV-SF galaxies
(particularly in the rest-frame UV) suggests that dustier galaxies have
star-formation which is less uniform, and/or that they suffer from more
structured dust obscuration.

Turning to the asymmetries, in both the rest-frame UV and optical, the
SMGs have comparable asymmetries as the high-$z$ field and UV-SF
samples ($A^{opt}({\rm SMG})=$0.27$\pm$0.03 and $A^{nir}({\rm
  SMG})=$0.25$\pm$0.02).  In the asymmetries alone, it therefore
appears that the SMGs are no more likely to appear as major mergers in
the rest-frame UV/optical than more typical (lower bolometric
luminosity), high-redshift galaxies.

\begin{figure*}
\centerline{
\psfig{file=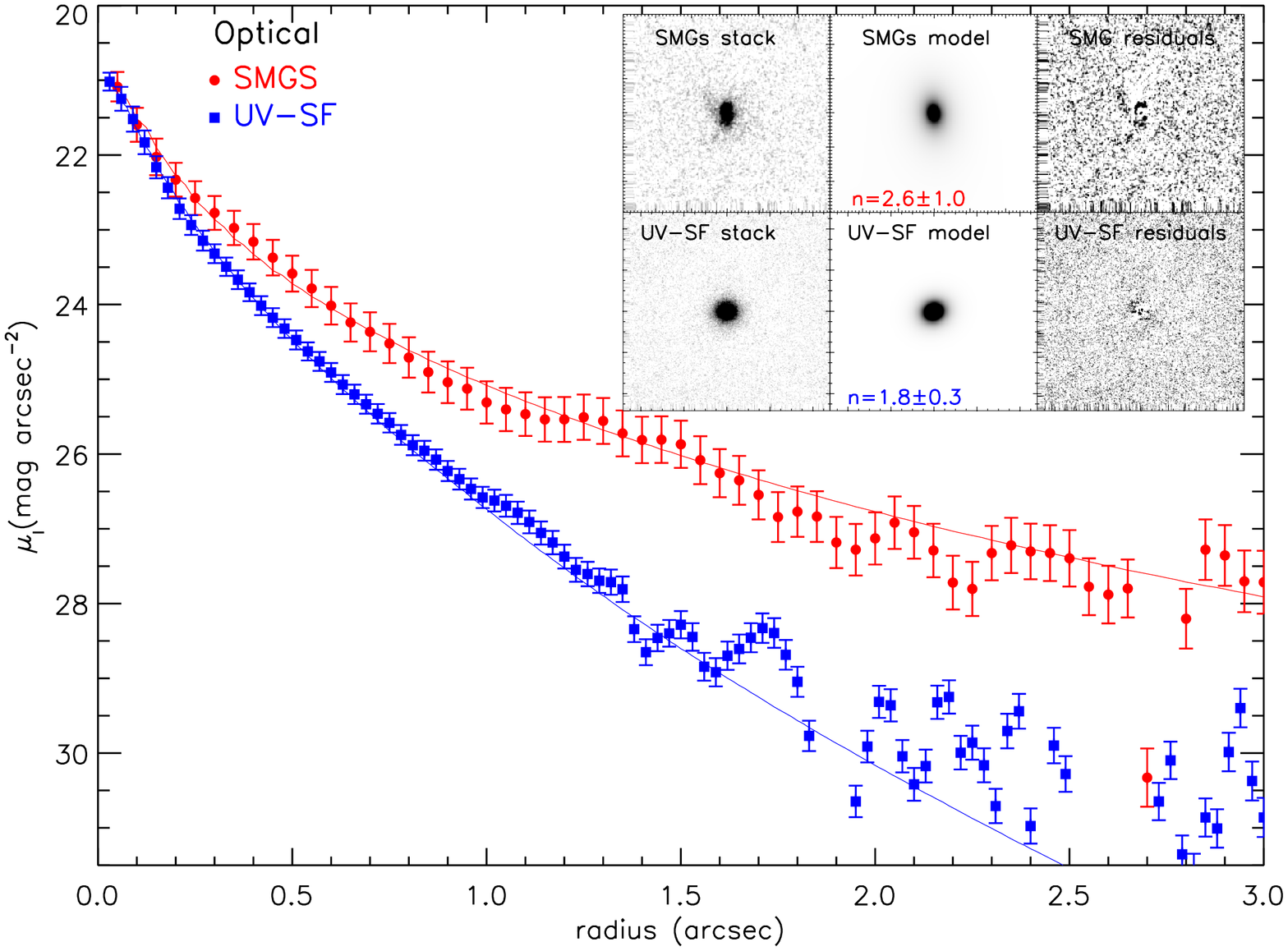,width=3.0in,angle=90}
\psfig{file=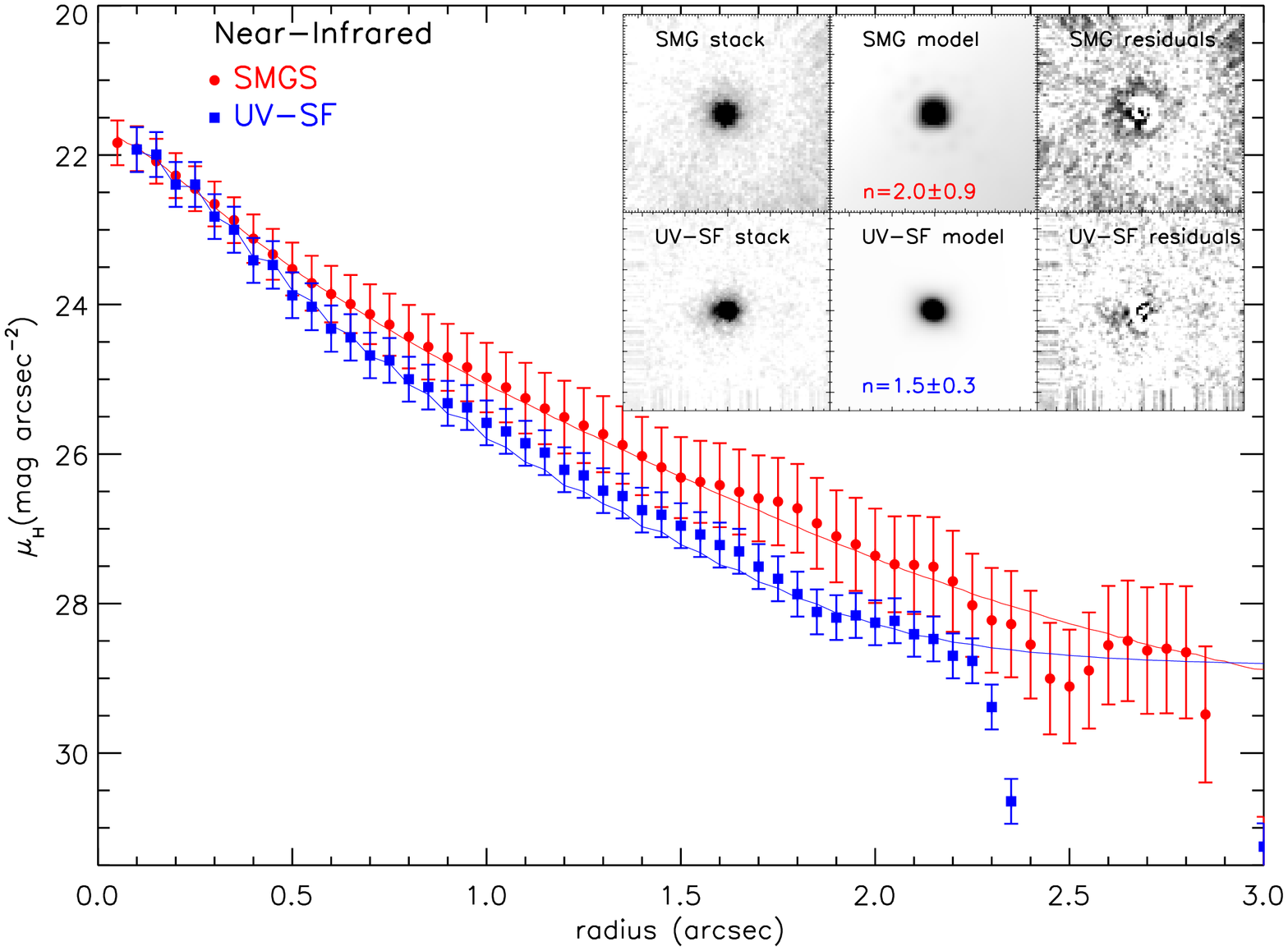,width=3.0in,angle=90}}
\caption{Average spatial light distribution of the SMGs and UV-SF
  galaxies in our sample in the observed $I$- and $H$-bands in the left
  and right hand panel respectively.  In both panels, the flux scale is
  arbitrarily normalised and the $I$- and $H$- band profiles are offset
  in flux scale for clarity.  The solid curves in both represent the
  best fit profiles.  {\it Inset:} the optical and near-infrared stacks
  of the galaxies with the best-fit model and residuals.  }
\label{fig:light_profile}
\end{figure*}

\subsection{Light Profile}

Given the high resolution and reasonable signal-to-noise of our data,
we also model the surface brightness distribution of the SMGs and
comparison samples.  We use {\sc galfit} \citep{Peng02} to model both
the $I$- and $H$-band surface brightness distributions for the sample.
To ensure robust results, we restrict the analysis to galaxies brighter
than I$_{AB}<$23.8 and for $H_{AB}<$24.5 (see \citealt{Cimatti08}).
For each galaxy image, the PSF which {\sc galfit} uses during the
fitting process was generated with {\sc tinytim}.  In all cases, we
used the Sersic profile to model the galaxy surface brightness profile,
allowing the Sersic index, $n$, to vary between $n$=0.5 and $n$=6.  For
$n$=1 and $n$=4 the Sersic profile reduces to an exponential or de
Vaucoulers profile respectively.  Disk dominated galaxies typically
have low Sersic indices ($n<$2), whilst bulge dominated galaxies tend
to have higher Sersic values (e.g. $n>$2).

Individually, the SMGs have a wide range of Sersic indices in both the
$I$- and $H$-bands, ranging from $n\sim$0.5 to $n$=4.5, with a median
$n_I$(SMG)=1.8$\pm$1.0 and $n_H$(SMG)=1.4$\pm$0.8 (the errors denote
the error on the mean; the errors on individual measurements are
typically $\pm$1.0).  Similarly, the UV/optical star-forming comparison
sample has a median Sersic index of $n_I$(UV-SF)=1.2$\pm$0.8 and
$n_H$(UV-SF)=1.5$\pm$0.7, although also with a comparably wide spread
as the SMGs.

An alternative test of the average light profile can be made by
stacking the individual images.  To achieve this, we first normalise
the light distribution using the total flux within the Petrosian
aperture (we only include galaxies brighter than I$<$23.8, $H<$24.5 as
above), and use the centre as defined by the Asymmetry minimisation
procedure.  In Fig.~\ref{fig:light_profile} we show the stacked
two-dimensional images, as well as the average light profiles for the
SMGs and comparison samples.  We also show the best-fit model and
residuals after subtraction of the best fit profile.  Using {\sc
  galfit} to model the light profile (accounting for the PSF using {\sc
  tinytim}) the resulting best-fit model in the $I$-band has
$n_I$(SMG)=2.6$\pm$0.5, whilst the $H$-band light profile is best
described with $n_H$(SMG)=2.0$\pm$0.5.  Although the SMGs are extended
on scales greater than the PSF in both $I$- and $H$-band (typically
r$_{h}>$0.25$''$;2\,kpc), it is likely that the spatial light profiles
are sensitive to the PSF.  We therefore convolve the $I$-band image
with the NICMOS PSF and the $H$-band image with the ACS PSF and re-fit
the $I$- and $H$-band images with the convolved PSF respectively,
obtaining $n_I$(SMG)=2.5$\pm$1.0 and $n_H$(SMG)=3.0$\pm$0.9.  Since
this extra smoothing produces Sersic indices within 1$\sigma$ of the
previous measurements, we conclude that the Sersic indices give a
reasonable estimate of the true Sersic indices of the galaxies.  We
also note that removing those SMGs with possible AGN contributions
(Table~1), the Sersic index increases in both cases by
$\Delta$n=0.2$\pm$0.3 (where the error accounts for the lower number of
galaxies in the stack).  Thus, the best fit SMGs in both $I$- and
$H$-bands is $n\sim2.0$--2.5 which is consistent with bulge-dominated
galaxies.  The same stacking procedure for the UV-SF comparison sample
yields a slightly smaller Sersic index with $n_I$(UV-SF)=1.8$\pm$0.3
and $n_H$(UV-SF)=1.5$\pm$0.3, significantly overlapping the Sersic
values in both bands for the SMGs.

\section{Discussion}

Using deep {\it HST} $I$ and $H$-band imaging, we have performed a
quantitative morphological analysis of a sample of 25 spectroscopically
confirmed sub-mm selected galaxies.
We measure the sizes of the SMGs, and derive typical half light radii
of r$^{opt}_h$(SMG)=2.3$\pm$0.3\,kpc 
and r$^{nir}_h$(SMG)=2.8$\pm$0.4\,kpc.
We find that the SMGs have comparable sizes to UV-SF galaxies, and the
general $z=1$--3 field population in both bands.  The sizes we derive
are slightly larger, but comparable to the gas sizes measured using
IRAM/PdBI millimeter interferometry ($r_{1/2}$=1.8$\pm$0.8\,kpc;
\citealt{Tacconi08}).  The similarity between the CO and observed
$H$-band (stellar) sizes suggest that the stars and dense gas are most
likely co-located.

Although the measured sizes of SMGs and UV-SF samples are comparable,
previous results have shown that there are significant differences in
stellar and dynamical masses, with SMGs having potential well depths
much greater than optically selected star-forming galaxies
\citep[e.g.][]{Swinbank04,Erb06a,Genzel06,ForsterSchreiber06,Swinbank06b,Law07}.
If the regions sampled by the H$\alpha$ and CO emission lines are the
same as those seen in the rest-frame UV/optical imaging, then this
suggests that the SMGs have surface matter densities up to an order of
magnitude larger than typical BX/BM and LBGs (see also
\citealt{Tacconi08}).

Using the asymmetry parameter to gauge the importance of major mergers,
we find that the SMGs and UV-SF comparison samples all have comparable
asymmetries in the optical and near-infrared, with $A=0.27\pm0.03$ and
$A=0.25\pm0.02$ in the observed $I$- and $H$-band respectively.  This is
somewhat larger than typically measured for spiral or elliptical
galaxies in the local Universe (typically $A<0.05$), but comparable to
the asymmetry measured in local ULIRGs ($A=0.35\pm0.1$;
\citealt{Conselice03a}).  Overall, (and surprisingly) this suggests
that in the the rest-frame UV/optical morphologies, SMGs are as equally
likely to appear as major mergers than lower-luminosity (and hence more
quiescent) high-redshift star-forming galaxies.  However, the Gini
co-efficients of the SMGs are systematically larger than the UV-SF
galaxies, suggesting less uniform star-formation in the rest-frame UV,
and possibly structured dust obscuration (see also \citealt{Law07b}).

The striking result that the SMGs have comparable sizes and asymmetries
as other high-redshift populations is at odds with the traditional
picture of SMGs which have shown that SMGs are extended starbursts
which are a result of major mergers
\citep[e.g.][]{Greve05,Swinbank06b,Tacconi08}.  However, many of these
studies have concentrated on other multi-wavelength data (such as
kinematic studies through CO or H$\alpha$).  Indeed, recently Ivison et
al. (2010) use deep EVLA radio imaging of the CO(1-0) to show that the
archetypal lensed SMG behind Abell\,1835 (L1L2) has a large, extended
cold molecular gas reservoir which extends over 25\,kpc in projection.
To reconcile this apparent contradiction, we measure the half light
radius, Gini co-efficient and asymmetry of L1L2 using the source-plane
$HST$ imaging, deriving half light radii of r$_h$=3.0$\pm$0.3 and
2.8$\pm$0.4\,kpc, asymmetries of $A$=0.32 and 0.30 and Gini
co-efficients of G=0.82 and 0.88 in observed $I$- and $H$-bands
respectively (centered on L1 in this system).  Thus, the rest-frame
UV/optical morphology of L1L2 appears similar to the typical SMGs
studies here, even though the kinematics and spatial extent of the gas
in this system is complex.  Our findings suggest that, whilst
multi-wavelength studies (such as those carried out via radio, CO or
H$\alpha$) suggest archetypal SMGs may mark the sites of complex,
merging systems, this is poorly reflected in their rest-frame
UV/optical morphologies (or, at least it is difficult to differentiate
SMGs from other more quiescent high-redshift populations via rest-frame
UV/optical morphologies alone).

\begin{figure*}
\centerline{
\psfig{file=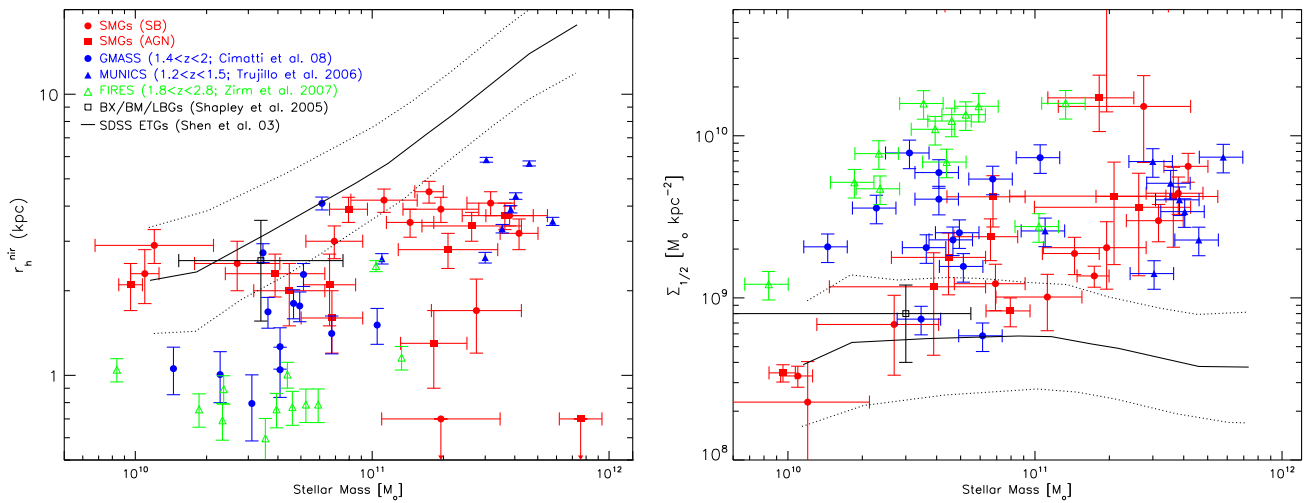,width=7in,angle=90}}
\caption{The distribution of physical sizes ($r_h$) versus stellar mass
  for SMGs compared to $K$-band selected samples of luminous red
  galaxies at $z\sim$1--2.  Stellar masses are estimates using the
  \citet{Maraston98} stellar population libraries.  The solid line
  shows the size-mass relation of early-type galaxies from SDSS by
  \citet{Shen03} (with dotted lines indicating the 1$\sigma$ scatter).
  The comparison samples comprise the GMASS passive, dense galaxies
  from \citet{Cimatti08}, passive galaxies selected from FIRES by
  \citet{Zirm07}, massive galaxies from MUNICS by \citet{Trujillo06}.
  We also include the optical/UV selected star-forming galaxy samples
  by combining their sizes (this work) and median stellar mass
  \citep{Shapley05}.  For clarity we split the SMGs into two samples:
  starburst (SB) and AGN where the classification is based on the
  spectroscopy of \citet{Chapman05a,Swinbank04,Takata06}.  This figure
  shows that approximately half the SMGs have comparably high stellar
  mass densities as the passive, luminous red galaxies from GMASS, but
  with substantially larger sizes and stellar masses (by a factor
  $\sim$2$\times$ in both cases). }
\label{fig:mass_size}
\end{figure*}

Finally, to test how the stellar densities of the SMGs compare to other
high redshift populations we combine the size and stellar mass
estimates and show the size versus stellar mass relation in
Fig.~\ref{fig:mass_size}.  We use the latest estimates for the SMG
stellar masses from Hainline et al.\ (2009a,b) (which take account of
the TP-AGB phase of stellar evolution; \citealt{Maraston98}) and AGN
contribution to the rest-frame near-infrared photometry.  For the SMGs,
the median stellar masses,
(M$_{\star}=1.4\pm0.5\times10^{11}$\,M$_{\odot}$) suggest stellar
densities which are a factor $\sim$5$\times$ larger (on average) than
the UV-SF population \citep[see also][]{Borys05,Alexander08}, but
slightly larger (by a factor 2$\times$) than local early type galaxies
\citep{Shen03}.  We also compare the sizes and stellar densities to
other $z\sim$1.5 luminous red galaxies (LRGs) from
\citet{Cimatti08,Zirm07} which (due to their moderately high stellar
masses, $\sim$10$^{10.5-11.0}$M$_{\odot}$, compact sizes
$r_h=1.2\pm0.2$\,kpc, and low space densities,
$\sim$10$^{-4}$\,Mpc$^{-3}$) have been proposed as direct SMG
descendants.  However, the stellar masses and sizes of the SMGs are, on
average, both a factor $\sim$2$\times$ larger than LRGs.  Thus, unless
the stars forming in the gas reservoir ultimately have a very different
spatial distribution than the gas itself (which is unlikely given the
similarity between the observed stellar and gas sizes), it is difficult
to reconcile the small physical sizes of luminous red galaxies at
$z\sim$1.5 with an evolutionary sequence with SMGs which already appear
to be a factor $\sim2\times$ larger measured at the same rest-frame
wavelength at $z\sim2.5$.

\section{Conclusions}

We have undertaken the first large near-infrared morphological
analysis of SMGs.  
We find that the SMGs have comparable sizes to UV-SF galaxies at the
same epoch (BX/BM and LBGs) in both $I$- and $H$-bands, and
(surprisingly) with comparable asymmetries.  However, we find that the
SMGs have systematically larger Gini co-efficients (particularly in the
observed $I$-band) than UV-SF galaxies at the same epoch suggesting
less uniform, high intensity star-formation in the rest-frame UV,
possibly reflecting structured dust obscuration (see also
\citealt{Law07b}).  
Overall our results suggest that SMGs are no more likely to appear as
major mergers in the rest-frame UV/optical than more typical,
lower-luminosity high-redshift galaxies (such as Lyman break galaxies
or BX/BMs).  However, the differences in the Gini coefficients between
populations suggests that the dustier SMGs have star-formation which is
less uniform (and/or that they suffer from more structured dust
obscuration).

We also show that most of the SMGs have observed $H$-band light
profiles which are better fit with that of a spheroid galaxy light
distribution.  Indeed, stacking the galaxies we find that in the
observed $H$-band, an n$\sim$2 Sersic index provides a better fit to
the spatial light profile than an exponential disk model, suggesting
that, whilst these galaxies are individually morphologically complex,
the composite stellar structure of the SMGs reflects that of a
spheroid/elliptical galaxy.  However, we note that the same analysis of
the UV-SF galaxies is statistically indistinguishable, with only a
marginally lower Sersic index with $n=1.6\pm0.3$.

The close similarity between the rest-frame UV and optical morphologies
in the SMGs, and UV-SF galaxies suggests that both wavelengths are
dominated by young, star-bursting components as well as dusty regions
\citep[see also][]{Dickinson00,Papovich05}.  These results are in
contrast to local studies of similarly luminous LIRGS and ULIRGs in the
local Universe which have been shown to have very different Hubble
types from the rest-frame UV and optical wavelengths
(e.g. \citealt{Goldader02}), possibly suggesting fundamental
differences between starbursts at $z=0$ and $z\sim2$.

Although the sizes and structural properties of the SMGs and UV-SF
star-forming galaxies are similar, previous work has shown that the
dynamical masses of SMGs are up to an order of magnitude larger than
UV-SF galaxies
\citep[e.g.][]{Erb03,Swinbank04,Erb06a,Genzel06,ForsterSchreiber06,Swinbank06b,Law07}..
This suggests that the intense star-formation within SMGs does not
represent an evolutionary sequence in which typical UV-SF galaxies
undergo intense star-formation (either through merging or through
secular processes), but rather it is the availability of large gas
reservoirs within already massive galaxies that allow the SMG phase.

Finally, we also investigate the size--stellar mass relation of SMGs in
order to test whether the SMGs may represent progenitors of the
luminous red galaxies seen at $z\sim$1.5.  We combine estimates of the
size and stellar masses to show that approximately half of the sample
have stellar mass densities comparable to those derived for luminous
red galaxies \citep[eg.][]{Cimatti08}.  However, we show that the
median size of the SMGs in the observed near-infrared
($r_{h}=2.3\pm0.3$\,kpc) is larger than that of the luminous red
galaxies, which have a median half light radius of
$r_{h}=1.2\pm0.2$\,kpc.  We also show that the median stellar mass of
the SMGs is also a factor $\sim2\times$ larger, thus suggesting that
the luminous red population at $z\sim1.5$ are unlikely to be direct
descendants of the SMG population unless the new stars formed in the
SMG starburst ultimately have a very different spatial distribution
from their gas reservoirs, which seems unlikely given the similarity
between the CO and UV/optical sizes.

Overall, our results suggest that rest-frame UV and optical
morphologies of high-redshift galaxies are essentially decoupled from
other observables (such as bolometric luminosity, stellar or dynamical
mass).  Alternatively, the physical processes occuring within the
galaxies are too complex to be simply characterised by the rest-frame
UV/optical morphologies.  It may be that at significantly longer
wavelength structural differences will appear, but this will have to
wait for rest-frame near-infrared imaging with {\it James Webb Space
  Telescope, (JWST)}.  Alternatively, high resolution kinematical
studies on sub-kpc scales (e.g. with near-infrared IFUs which can now
be carried out from the ground using adaptive optics) may offer the
most direct route to probing the differences between high-redshift
galaxy population as the dynamics, distribution of star-formation and
metallicity gradients will reflect differences in the triggering
mechanism and mode of star-formation at high-redshift.

\section*{acknowledgments} 

We are very grateful to the referee for their constructive comments
which significantly improved the content and clarity of this paper.  We
would like to thank Alfred Schultz at STScI for advice on dealing with
the effects of the SAA on our NICMOS data.  We gratefully acknowledge
Eric Richards for providing us with his reduced maps of HDF and SSA13
and David Law and Jim Dunlop for useful discussions.  AMS gratefully
acknowledges a Sir Norman Lockyer Royal Astronomical Society
fellowship.  SCC acknowledges support from NASA grants \#9174 and
\#9856.  IRS acknowledges support from STFC.  AWB ackowledges NSF grant
AST-0205937 and the Alfred P.\ Sloan Foundation.  DMA thanks the Royal
Society and Leverhume trust.

\bibliographystyle{apj}
\bibliography{/Users/ams/Projects/ref}
\bsp

\end{document}